\documentstyle[12pt,aasms4]{article}

\def\T1{\ {$T_1$}\ }
\def\CT1{\ {$(C-T_1)$}\ }

\def\Wash{\ {Washington}\ }
\def\met{\ {metallicity}\ }
\def\mets{\ {metallicities}\ }
\def\cl{\ {cluster}\ }
\def\cls{\ {clusters}\ }
\def\mp{\ {metal-poor}\ }
\def\mr{\ {metal-rich}\ }
\def\redd{\ {reddening}\ }

\def\phot{\ {photometry}\ }
\def\pop{\ {population}\ }
\def\pops{\ {populations}\ }
\def\magn{\ {magnitude}\ }

\def\ell{\ {elliptical}\ }

\def\gtsim{\ {\raise-0.5ex\hbox{$\buildrel>\over\sim$}}\ }
\def\ltsim{\ {\raise-0.5ex\hbox{$\buildrel<\over\sim$}}\ }

\begin{document}

\title{Ages and Metallicities of Star Clusters and Surrounding Fields \\
in the Outer Disk of the Large Magellanic Cloud}

\author{Eduardo Bica}
\affil{Departamento de Astronomia, Instituto de F\'{\i}sica, UFRGS,
        C.P. 15051, 91501-970  \\
Porto Alegre RS, Brazil}

\author{Doug Geisler}
\affil{Kitt Peak National Observatory, National Optical Astronomy Observatories,\\
P.O. Box 26732, Tucson, Arizona 85726}

\author{Horacio Dottori}
\affil{Departamento de Astronomia, Instituto de F\'{\i}sica, UFRGS,
        C.P. 15051, 91501-970  \\
Porto Alegre RS, Brazil}

\author{Juan J. Clari\'a\altaffilmark{1}, Andr\'es E. Piatti\altaffilmark{1}}
\affil{Observatorio Astron\'omico de C\'ordoba, Laprida 854, 5000, \\ C\'ordoba,
Argentina}

\author{and}

\author{Jo\~ao  F. C. Santos Jr.}
\affil{Dep. de Fisica, ICEx, UFMG,
C.P. 702, 30123-970 Belo Horizonte MG, Brazil}

\slugcomment{Submitted to the Astronomical Journal}

\altaffiltext{1}{ Visiting Astronomer, Cerro Tololo Inter-American
Observatory, which is operated by AURA, Inc., under cooperative
agreement with the NSF.}

\begin{abstract}
We present Washington system $C,T_1$ color-magnitude diagrams of 13 
star clusters and their surrounding fields which lie in the outer parts of the LMC
disk ($r>4\arcdeg$), as well as a comparison inner 
cluster. 
The total area covered is large
($2/3^{\Box^\circ}$),
allowing us to study the clusters and their fields individually 
and in the context of the entire galaxy.
Ages  are determined by means of the magnitude difference $\delta$T$_1$ between
the giant branch clump and the turnoff, while metallicities are derived from the
location of the giant
and subgiant branches as compared to fiducial star  clusters. This
yields a unique dataset in which ages and metallicities for both a significant
sample of clusters and their fields are determined homogeneously. We find 
that in most cases the stellar population of each star cluster
is quite similar to that of the field 
where it is embedded, thus sharing its mean  age and metallicity. 
The old population (t$\geq$10 Gyr) is detected in most fields as a small  concentration
of stars on the horizontal branch blueward and faintward of the prominent clump. 
Three particular fields present remarkable properties: (i) The so far unique cluster
ESO121-SC03 at $\approx$9 Gyr has a  surrounding field which shares  the same properties, 
which in turn is also unique in the sense that such a dominant old field component  is not 
present elsewhere, at least  not significantly in the fields as yet  studied. 
(ii) The field surrounding the far eastern 
intermediate age cluster OHSC\,37   is noteworthy in the sense 
that we do not detect any evidence of LMC stars: it is 
essentially a Galactic foreground field. We can thus detect the LMC field out to $>11\arcdeg$ 
(the deprojected distance of ESO121SC03), or $\sim 11$ kpc, but not to $13\arcdeg (\sim 13$
kpc), despite the presence of \cls at this distance. (iii) In the northern part of the LMC disk
the fields of SL388 and SL509 present color-magnitude
diagrams with a secondary  clump  $\approx$0.45 mag fainter than the dominant 
intermediate age clump, suggesting a stellar population 
component    located behind the LMC disk at a distance 
comparable to that of the SMC. 
Possibly we are witnessing a depth effect in the LMC, and the size of the 
corresponding structure
is comparable to the size of a dwarf galaxy.  The unusual spatial 
location of the cluster OHSC37 and the anomalous properties 
of the SL\,388 and SL\,509 fields  might be explained as  debris  from   previous LMC 
interactions with  the Galaxy and/or the SMC.

The mean    metallicity derived for the intermediate age outer disk clusters 
is $<$[Fe/H]$>$$=-0.7$ and for their surrounding fields 
$<$[Fe/H]$>$$=-0.6$. These values are significantly lower than found by
Olszewski {\it et al.} (1991, AJ, 101, 515) for a sample of clusters of similar age, but are
in good agreement with several recent studies.
A few clusters stand out in the age--metallicity relation in the sense that they are 
intermediate age clusters at relatively low \met ([Fe/H]$\approx -1$).
 
\end{abstract}
\eject

\section{INTRODUCTION}

Unveiling the   star formation  history and  chemical enrichment of galaxies is critical 
for understanding how they form and 
evolve. In this respect, Local Group galaxies play a fundamental 
r\^ole (Hodge 1989). The  proximity of the Large Magellanic Cloud allows one to probe 
its stellar population properties with different techniques, namely photometry and 
spectroscopy of individual stars in clusters and the field, and integrated methods  
in the case of  star clusters. Such studies for nearby galaxies are important to better 
understand very distant galaxies, whose stellar populations can only be probed   by means 
of integrated properties. 

     Concerning field color-magnitude diagram (CMD) studies, ground-based observations 
have allowed the accurate  
study of the brighter evolutionary sequences, sampling relatively large fields
throughout the LMC bar and disk (e.g. Butcher 1977, Hardy  {\it et al.} 1984, Bertelli {\it et al.} 1992, 
Westerlund, Linde \&  Lyng\aa\ 1995, Vallenari {\it et al.} 1996). A prominent intermediate age (1-3 Gyr) 
stellar population is universally
present in  these fields, together with varying amounts of  young blue  
main sequence (MS) stars.   Hubble Space Telescope (HST) observations are limited
to a small 
viewing area,  but in turn allow  deeper photometry, to well below the old MS turnoff. 
Several such fields have been studied with  the V and I bands. One, at $\approx$4$^o$ 
north of the bar (Gallagher {\it et al.} 1996) and another near the SE end of the bar (Elson, 
Gilmore \& Santiago 1997) show evidence for the major star-forming event to have ocurred
$\approx$2 Gyr ago, in agreement with the ground-based studies.
More recent HST studies (Holtzman {\it et al.} 1997, Geha {\it et al.} 1998) have investigated
three LMC fields at 3$^o$ to 4$^o$ from the bar center, one 
located in the  north-east and two in the north-west. Surprisingly, they are finding
many more faint MS stars than expected and suggest that there has been more star formation
in the past than previously believed. Their models, assuming a standard IMF slope, suggest
that fully one half of the stars in these fields were formed more than 4 Gyr ago.
Although the latter studies refer to their fields as ``outer", 
we point out that in the present work we deal with genuine outer disk fields, well
beyond any of these HST studies. 

    Integrated photometry of large star cluster samples of all ages have shown differences 
in the spatial distribution of age groups  both in the bar region (Bica, Clari\'a \& 
Dottori 1992) and the entire LMC (Bica {\it et al.} 1996). Differences in the  spatial distribution 
among  young groups have provided insight  on the formation process  and subsequent dynamical 
evolution of star cluster generations (Dottori {\it et al.} 1996). This integrated photometry cluster
sample has been compared  with integrated star cluster color models and has provided 
constraints on the cluster formation history (Girardi \& Bica 1993, Girardi {\it et al.} 1995).

  CMDs of LMC star clusters have also revealed a large intermediate 
age population  (1-3 Gyr), which is separated  by a pronounced age gap  from the old 
stellar population as denoted by  a few genuine globular clusters (see Da Costa 1991, 
Suntzeff {\it et al.} 1992, Olszewski, Suntzeff \& Mateo 1996 for reviews). Recently, CMDs 
in the  Washington system of a sample of candidate old clusters selected from the Bica {\it et al.}
(1996) and Olszewski {\it et al.} (1991) studies revealed 
them to instead be of intermediate age  (Geisler {\it et al.} 1997, hereafter Paper I).
This study increased  considerably the known sample of 1-3 
Gyr old clusters with accurate age determinations,  and reinforced 
the conclusion that a major formation epoch was preceded by
a quiescent  period of many Gyr, or that cluster dissipation has been more effective than
generally believed (e.g. Olszewski 1993).
   
   The objective of the present paper is to compare the properties of outer LMC clusters 
with those of their surrounding fields  by using  the same observational technique,  and 
to infer  the age-metallicity relation, and whether it depends  on the  spatial distribution 
throughout the LMC disk. In order to achieve this we employ Washington system  C, T$_1$  
bands and construct CMDs using the data of Paper I.
Ages are inferred from the difference $\delta$T$_1$ between 
the giant branch clump and the turnoff (see also Paper I), and from  the occurrence  of 
particular stellar evolutionary sequences in the CMDs.   The giant and subgiant branches 
allow one to derive metallicities using a technique analogous to that of Da Costa and 
Armandroff (1990) for VI photometry. However, our combination of Washington system filters is 
three times   more \met sensitive than the VI system (Geisler \& Sarajedini 1996, 1998),
allowing us to obtain accurate \mets for both the clusters and their fields. The 
cluster/field sample and the observations are described in Section 2. The  cluster and field 
CMDs are described in Section 3. Ages and metallicities
are derived in Section 4. In Section 5 we discuss  the chemical enrichment of the outer disk,
and in Section 6 the presence of dual clumps in two fields is noted and the 
possibility of a depth effect in
this portion of the LMC disk is discussed.
Finally, the conclusions of 
this work are given in Section 7.

\section{The Sample and Observations}

The outer LMC disk is tilted at  i$\approx$45$^o$ to the line of sight with the line of 
nodes at $\Theta\approx$7$^o$,  as indicated by the distribution of outlying star clusters 
(Lyng\aa\ \& Westerlund 1963, see also Westerlund 1990 for a review). The total observed sample
of clusters was first reported in Paper I. In the present paper we discuss in detail  
the 13 most distant
outlying clusters  from this sample and an additional inner cluster, SL769, included for
comparison purposes, and their respective 13 surrounding fields 
(IC2134 and SL451 are located in the same frame).  The star cluster designations in different 
catalogs and respective  1950 equatorial coordinates are listed in Table 1. Galactic coordinates 
are also given in the Table: since the distribution of the present clusters covers as much as 
15$^o$ on the sky, variations of Galactic reddening are expected.

      As a rule the clusters are centered in the frames, with some exceptions which were shifted
to avoid bright stars. In the case of IC2134 and SL451 the frame
was centered approximately halfway between the two clusters, as illustrated in Figure 1, which
also serves to illustrate typical clusters and field. The spatial 
distribution of the clusters is shown in Figure 2. With the exception of OHSC37 which is located far 
east away from the   disk  body,  the remaining clusters are consistent with an inclined disk 
as described above. We also show in Table 1  the approximate projected angular distance 
from the bar center (taken  as the position of the cluster NGC1928  
$\alpha_{1950}$ = 5$^h$ 21$^m$ 19$^s$, $\delta_{1950}$ =  -69$^o$ 31' 30''), which in turn 
is $\approx$0.2$^o$  south of   the HI rotation curve center (see Westerlund 1990 for a review of 
centroids). Finally, the last column of Table 1 lists deprojected distances R assuming that 
all clusters are part of the inclined disk.

    The observations were carried out with the CTIO 0.9m telescope  in December, 1996
with the Tek2k \#3 CCD, as described in Paper I.
The scale on the chip is $0.40\arcsec$ per pixel, yielding an area $13.6\arcmin\times13.6  
\arcmin.$
SL769 was observed with the CTIO 4m in February, 1996 with the Tek2k \#4 CCD, with similar 
pixel and areal coverage. The filters used for both runs were the Washington (Canterna 1976) 
C and Kron-Cousins R filters.  The latter has significant throughput
advantages over the standard \Wash T$_1$ 
filter (Geisler 1996). In the present work, as in Paper I, we calibrate  the observations
in the  C, T$_1$ system. In particular this filter combination allows us  to derive accurate 
metallicities based on the standard giant branch technique outlined in Geisler \& Sarajedini 
(1996, 1998). The data were reduced with the stand-alone version of DAOPHOT II (Stetson 1987) 
after trimming, bias subtraction and flat-fielding. More details on the observations,
reductions and calibration procedures were given in Paper I.
 
After deriving the \phot for all detected objects in each filter, a cut was made on the 
basis of the parameters returned by DAOPHOT. Only objects with $\chi <2$, 
photometric error $<2\sigma$ more than the mean error at a given magnitude, and
$\mid Sharp \mid <0.5$ were kept in each filter (typically discarding about $10\%$ of the
objects), and then the remaining objects in the C and \T1
lists were matched with a tolerance of 1 pixel, and raw \phot obtained. This raw \phot
was then transformed to the standard \Wash system as outlined in Paper I. Final calibrated
\phot for each \cl (Table 2a-n) is available on the electronic AJ database.

We  present in Figs. 3a to 3n the star cluster CMDs.
The cluster radii were selected by eye judging the variation of the stellar 
density in the cluster surroundings, and ranged from 50-200 pixels, with a typical
value of 75 pixels ($30\arcsec$). These radii are intended to optimize the cluster
CMD sequences to avoid  field contamination as much as possible. The radius in pixels 
for  the extraction of  each cluster  is indicated in the corresponding CMD in Fig. 3.

We checked on the extent of field contamination of each \cl CMD by obtaining an equal-area
field CMD composed of the addition of CMDs derived from 4 different fields, each of an area 
1/4 that of the \cl and lying far away from the \cl. Note that such a comparison will
overestimate field contamination since the photometric limit within the \cl will be brighter
than in the field and more stars will be discarded from the \cl because of larger photometric
errors, due to increased crowding. Nevertheless, in only 1 outlying \cl was the number of 
stars obtained in the equal-area field significantly more than 1/4 of the stars in the \cl
CMD. The typical ratio was only $\sim 10\%$. Thus, we did not perform any statistical 
subtraction of the field contamination from the \cl CMDs.

We have also checked cluster extent on the basis of star counts and structural parameters 
and how it may be contributing to the surrounding fields in our frames. Three clusters in our
sample have been studied by Kontizas, Hatzidimitriou \& Kontizas  (1987), namely SL\,388 (LW\,186), 
IC\,2134 (LW\,198) and SL\,842 (LW\,399) for which they could detect cluster stars as
far as $r=1.4'$, $r=1.4'$ and $r=1.7'$ respectively.
Our selected radii for these \cls were $0.5\arcmin, 0.5\arcmin$ and $0.4\arcmin$. Thus, our
\cl radii are conservative and should minimize field contamination.

We have also obtained CMDs for each field, 
excluding the pixels within a radius twice  that used
for the cluster, which minimizes contamination of the field by
cluster stars.
The tidal radii derived by Kontizas {\it et al.} for the three clusters ($r_t=3'$,
$r_t=6.2'$ and $r_t=3.1'$ respectively) are  within the limits of the fields so defined.
However, our criterion of field extraction is
comparable to the detection limits of star counts, and
the contribution of \cl stars to the field CMD,
especially from \cls as relatively poorly populated
as our sample, should not be significant.
The tidal radius of ESO121-SC03 is  $2.35\arcmin$ according to Mateo,
Hodge \& Schommer (1986). Notice that the cluster CMD extraction (Figure 3i)
is for r$<1.33\arcmin$, so that the field extraction corresponds to r$>2.67
\arcmin$,
thus beyond the cluster tidal radius.
The field CMDs are shown in  Figs. 4a to 4m.

\section{Description of  CMDs}

The cluster CMDs (Fig. 3) are typical of intermediate age clusters (IACs) with turnoffs ranging
from \magn levels slightly below the clump of He-burning stars to as much as 2 mag. below.
The only exception is ESO121SC03 which is considerably older. See Paper I for a discussion
of the clump to turnoff magnitude difference and its relation to age.

\subsection{Fields}
    The uniform large size of the present frames is particularly suitable for the study of the brighter sequences of field stellar populations.
The number of stars seen in the field CMDs (Fig. 4) is clearly  correlated 
with the deprojected angular distances R from the bar center (Table 1). The radial 
dependance of  several CMD features is also noteworthy. The field of OHSC37 at a 
distance  R=13$^o$ has no evidence of a clump or horizontal branch or turnoff of any kind,
so that no  
LMC field stellar  
populations are detected in this frame. 
The wide distribution of stars in the diagram 
essentially consist of  foreground  Galactic stars.  
Note that OHSC37 is the most distant LMC \cl in the Olszewski {\it et al.} (1988) outer LMC \cl
catalog. On the other hand,  the field of ESO121SC03 presents 
clear clump and turnoff regions at the LMC distance. 
The $\delta$T$_1\approx2.9$ value is 
very similar to that of the cluster  ESO121SC03 itself (Paper I), thus both cluster and 
LMC field have comparable ages ($\sim 9$ Gyr) at this far north locus with R=$11.4^o$.
No evidence exists for any intermediate age \pop in this field.
One possibility is that the cluster 
and  its surrounding field were  a building block of the LMC, such as an accreted dwarf companion. 
Note that Galactic contamination 
is still important in this field. 

Thus, we can detect the LMC field star \pop out to $>11\arcdeg$ (the deprojected distance of
ESO121SC03), or $\sim 11$ kpc, but not to $13\arcdeg (\sim 13$ kpc), despite the presence of \cls
at this distance. This is in good agreement with previous estimates of the extent of the LMC
and its \cl system (e.g. de Vaucouleurs 1955, Lyng\aa\ and Westerlund 1963, Olszewski {\it et al.}
1988).

The LMC becomes prominent relative to the Galactic field   at distances  
10$^o<$R$<$7$^o$ (e.g. the  SL126 and SL842 fields), and the principal turnoff is still well 
below the clump ($\approx$2 mag).  The oldest population is detectable by means of a 
concentration of stars  near  the instability strip locus faintward and blueward of the 
intermediate age clump. 

       For  less distant fields the younger MS turnoff rapidly 
brightens, reaching 
the clump magnitude level for fields at R$\approx$5$^o$ (e.g. that of  SL8). The 
composite turnoff structure is obvious for such fields. Finally,  the field 
of SL769 at R=4$^o$ definitely has an important component with a blue  MS brighter 
than that of the clump. We adopt this criterion as a definition of the inner LMC disk.
In this inner disk field, a minor old component, as demonstrated
by the HB stars faintward and blueward of the clump, is still present.
The properties of our inner disk \cl sample and their fields
will be the subject of a forthcoming paper.

\section{Derivation of Ages and Metallicities}

\subsection {Ages}

The utility    of age determinations based on the magnitude 
difference between the clump/HB and the turnoff for IACs             and old clusters
is well known (see Phelps, Janes \& Montgomery  1994 for Galactic open clusters). 
   In Paper I we defined and calibrated such a method for   
$\delta$T$_1$ and applied it to our LMC cluster sample, including most of the present 
ones. In the present study we have rigorously eliminated photometric outliers.
As a consequence, the CMDs were improved. We remeasured $\delta$T$_1$ values and 
in some cases small differences appeared with respect to Paper I values. 
Such differences were almost always within the errors, averaging only 0.15 Gyr.
Age determinations by two independent investigators yielded ages within 0.3 Gyr in the mean,
with a standard deviation of 0.3 Gyr.
We also ranked the clusters according to
turnoff morphology and reddening corrected magnitudes taking into account the 
foreground E(B-V)$_G$ values (Table 4). The derived ages are given in Table 3.
These values should be preferred over those given in Paper I.

 Since the fields are in most 
cases obviously composite in age, we measure the  $\delta$T$_1$ value of the 
youngest well-populated turnoff and use the age calibration from Paper I to derive the 
representative age 
given in Table 3. Although there are some fields with small numbers of younger stars, 
most of the stars in these fields are as old as or older than the given age.
We cannot place any useful constraints on the proportion of turnoff stars in each field
older than $\sim 4$ Gyr as our \phot is not sufficient for the challenge. However, we do note
that in the field with our deepest photometry (SL769, obtained with the 4m)
one can follow the subgiant branch down to a level which corresponds roughly
with that shown by ESO121-SC03, i.e. the faintest subgiants in the SL769 field
are consistent with an old ($\sim 9$Gyr) age. In some of the other fields, similarly
faint subgiants can be seen.

Comparing the ages for the clusters with the surrounding fields as determined above
we conclude that they are similar, which suggests that in general the cluster has the same
origin as the surrounding field population. 

\subsection {Metallicities}

Da Costa and Armandroff (1990)
showed the utility of the $(V-I)$ color of the
red giant branch  for measuring metallicities in old stellar populations. 
This method
now sees very wide use, and is the preferred technique, e.g.,
for HST WFPC2 observations
of the stellar populations in distant Galactic globular clusters and nearby galaxies.

Geisler (1994) and Geisler and Sarajedini (1996)
introduced a similar technique using the $(C-T_1)$
color of the 
Washington system and demonstrated that it had much  potential for deriving
metal abundances in distant objects, with a metallicity sensitivity greatly exceeding
that of $(V-I)$. Geisler and Sarajedini (1998) have now derived the calibration of the 
standard giant branches in the Washington system technique. They have used 
the  mean loci of giant and
subgiant branches of Galactic globular and several old open 
clusters  with known metallicities  as fiducial \cls to derive the empirical relation
between the \CT1 color of the giant branch and \met, and show that this technique has
three times the sensitivity to \met that the corresponding $(V-I)$ technique has.
The ability to derive accurate \mets for our program \cls and fields using this new technique
was one of our primary motivations for using the \Wash system in the current study.

However, the fiducial \cls used by Geisler and Sarajedini (1998) are all either typical
Galactic globular \cls, with ages of $>10$ Gyr, or among the oldest open \cls (M67, with an
age of $\sim 4$ Gyr and NGC 6791, with an age of $\sim 10$ Gyr). In contrast, the only 
comparably aged \cl in the current sample is ESO121SC03 -- all of the rest are IACs        
ranging from $\sim 1-2$ Gyr. Given the noticeable effect of such a large age difference
on broadband colors, the Geisler and Sarajedini calibration is not directly applicable to
our sample. Instead, what we have chosen to do is to use the Geisler and Sarajedini 
fiducial loci to derive \mets for a sample of LMC and Galactic open \cls of intermediate
age and well-determined \met and determine whether any offset is found between the known
\met and the derived \Wash value. 

There are a total of five Galactic open \cls (Tombaugh 2, Melotte 71, NGC 2204, NGC 2506
and Melotte 66) and six LMC \cls (SL262, SL388, SL842, NGC 2213, OHSC33 and OHSC37) which
have both good \Wash \phot available  for a number of stars along the upper giant branch
(either this paper, previous publications or unpublished)
and accurate \mets (for the Galactic open \cls, these were taken from a variety of sources;
for the LMC \cls , \mets were taken from Olszewski {\it et al.} 1991). 
The ages of these \cls range from 1--4 Gyr. 

In order to compare these
\cls to the fiducial ones, a \redd and distance modulus were required to put the comparison
\cls together with the standard ones
in the $(M_{T_1},(C-T_1)_0)$ plane. Again, for the Galactic open \cls we used the best 
current estimates available for these values based on an extensive literature search. For
the LMC \cls, we  assumed a true distance modulus (m-M)$_0$=18.5 taking into account
results obtained from SN\,1987A ((m-M)$_0$=18.5, Panagia {\it et al.} 1991)
and the consequences of a recent revision of the Cepheid distance calibration
on the LMC distance modulus((m-M)$_0=18.50\pm0.15$, Madore \& Freedman 1998).
The extent of the LMC outer disk is considerable, so we decided to use a
foreground reddening E(B-V)$_G$ depending         on the Galactic coordinates
(Table 1) and the values from the maps by Burstein \& Heiles (1982).
The results are in Table 4. The reddening values for our entire LMC \cl sample
vary from 0.00 to 0.10, except for
OHSC37, which is located at a lower Galactic latitude and consequently has
a higher reddening.
Since we are dealing with the outer LMC, the disk internal reddening is expected
to be negligible. We note that an increase of the assumed reddening by E(B-V)=0.03
decreases the derived metallicity (see below) by 0.12 dex.

We then derived \mets for each of the 11 comparison \cls by interpolating by eye among the 
standard giant branches. Figure 5 shows the derived \Wash \met vs. the standard value
taken from the literature. A clear trend is found, indicating that the derived \Wash \mets
for IACs using this technique require an approximately constant zero point correction.
An unweighted mean yields a difference
of $0.46\pm0.21$ dex. We then determined the \met of all of our program LMC \cls 
and fields in the
same manner, and applied an offset of 0.45 dex to the derived value, i.e. we increased
our estimate derived from a direct comparison with the standard clusters by 0.45 dex.
Figure 6 shows a typical IAC     and Figure 7 a typical IAC field.
The final corrected values are given in Table 4.
In a few cases, the \mets were difficult or virtually impossible to determine because of
the lack of bright giants -- these cases are marked with colons or dashes in the Table
and any derived \mets are more uncertain. Note that the fields generally showed a 
significant range in metallicity, amounting to $\sim 0.5$dex (although some of this scatter
can be explained by Galactic field star contamination or LMC AGB stars), and that the
values quoted are crude means.
The offset is required for any intermediate age objects, which is the case for all but one
of our LMC \cls and most of the field stars. However, note that our value for 
ESO121SC03, as derived from Figure 8,
has not been corrected, since this is an old \cl. Our derived \met is only 0.17 dex lower
than that of Olszewski {\it et al.} for this \cl , showing good agreement.
In the case of any field stars that are significantly older than $\sim 4$ Gyr, our corrected
\met will represent an overestimate. However, such stars appear to be
a very small minority in our
fields.

In order to get an estimate of the errors involved in the metallicity derived from our
interpolation procedure,
two of us made independent measurements.  We got the following differences:  for metallicities of clusters and fields
 $<$[Fe/H]$_{EB}$-[Fe/H]$_{DG}$$>=-0.05$ with $\sigma$=0.07; for the metallicity 
difference between cluster and respective field  
$<$$\Delta$[Fe/H]$_{EB}$-$\Delta$[Fe/H]$_{DG}$$>=-0.07$ with 
$\sigma$=0.13 for 12 field/cluster CMDs. 
Therefore, we estimate that our internal \met errors are of the order of 0.1 dex and that
the total \met uncertainties are $\sim 0.2$ dex.

Table 4 also lists the Olszewski {\it et al.} \met values for \cls in common. For these 7 \cls, we
derive a mean difference of only 0.04 dex (our values are more \mr ), with $\sigma=0.34$ dex.
Given that their errors are similar to ours, this is exactly the value expected if no other
sources of error are present. Of course, we have used 5 of these clusters to derive the 
offset, but these represented $<$1/2 the total number of \cls used.

\section{Chemical Enrichment}

Olszewski {\it et al.} (1991) studied the chemical enrichment of the LMC based on
ages from CMDs and metallicities derived from  Ca\,II triplet  spectroscopy of  
some individual  giants. They obtained $<$[Fe/H]$>=-0.42$ for 17 \cls with ages in the 
same range as for our sample (1--3 Gyr).
In Paper I we increased the sample of well-studied IACs by
combining ages from new CMDs with metallicities from Olszewski {\it et al.} (1991).
We found evidence of a larger dispersion of metallicities at intermediate ages in the
sense that some outer clusters were more metal poor than the average. In the present study
we revisit this issue for our outer disk sample and their fields, using the ages  
and metallicities derived here.
Our method  relies  on a larger number of
stars per cluster (Olszewski {\it et al.} generally observed only $\sim 2$ stars per \cl while 
we generally have 5 or more to determine the giant branch locus); the relative \met errors
are similar. 

The results are shown in Figure 9, where we plot clusters and fields,
as well as some LMC globular clusters for comparison purposes (Paper I and references therein), and the mean curve for 
IACs derived by Olszewski {\it et al.}
The resulting enrichment scenario we find
for the outer disk is that of a pre-enriched gas with metallicity about
one tenth solar (the metallicity level of ESO121SC03 in the quiescent epoch). 
Most parts of the outer disk were enriched by $\approx$0.3  dex before or during  the 
burst which formed the IACs,                  
but some regions  apparently remained at the pre-enrichment level, forming some clusters with 
$[Fe/H]$$\approx-1.0$ even at later stages of the burst. 
Alternatively, such clusters might have been formed elsewhere and in their orbits are presently
superimposed or embedded in slightly more metal rich outer disk fields (see discussions in 
Paper I). Indeed, Fig. 9 reveals a considerable range  of 
metallicities for the IACs                  ($-0.5>$[Fe/H]$>-1.05$),
whereas the fields cover only 1/2 this range ($-0.4>$[Fe/H]$>-0.75$). The clusters 
responsible for the metallicity dispersion are SL509, SL862 and OHSC33, which have
$\approx$0.3 dex lower metallicities than the respective fields. 
Note that SL509 is not only metal-poor but also very young and that it exhibits the
dual clump discussed in the next section. OHSC33 is also intriguing as another young, \mp \cl.
The \met we derive for it is virtually identical to that found by Olszewski {\it et al.}
However, for most
of the sample, clusters and fields have comparable properties.
The mean metallicity of the 11 IACs                  
in the outer disk is $<$[Fe/H]$>$$=-0.71\pm0.17$,
while for the
fields it is $<$[Fe/H]$>$$-0.61\pm0.11$. 
A direct comparison shows that the \cls are essentially all
slightly more \mp than their respective
fields, with a mean difference of 0.12 dex. Thus, the \cls and their fields have very 
similar \mets.

The IACs in the present sample are on the average 0.29 dex more metal poor than 
\cls of similar age in the 
Olszewski {\it et al.} (1991) sample. We have already shown that \cls in common   
yield similar metallicities.
Probably our lower average  metallicity is real and reflects a gradient between the
outer and inner regions of the LMC (see also Paper I).
Figure 10 plots \met vs. deprojected radius R for both the \cls and fields in our sample.
No strong trend emerges but, as noted by Olszewski {\it et al.}, there is a tendency for the most
distant \cls to be more \mp. Note that Olszewski {\it et al.} did uncover $\sim 10$ IACs         
with \mets $<-0.75$ but their ages were mostly unavailable so they were not 
included in their age-metallicity relation.

Traditionally, the young ($<1$Gyr) LMC \pop has been found to have a \met only slightly
($\sim 0.2$ dex) less than solar (e.g. Harris 1983, Russell and Bessell 1989). IACs        
are generally regarded as having only slightly lower \mets ([Fe/H]$\sim -0.4$),
mostly based on the work of Olszewski {\it et al.} However, a number of studies have suggested
lower \mets for such objects might be more appropriate.
In an early study  including a large sample of IACs using narrow band integrated photometry,
Bica, Dottori \& Pastoriza (1986) derived metallicities similar  to the present ones.  
Although ages for LMC IACs in that study were mostly based on LMC calibrators available at that
time and require  revision, the metallicities were derived from reference Galactic open and 
globular clusters with accurate abundances.
Richtler {\it et al.} (1989) and Richtler (1993) derive values of --0.7 to --0.9
for several young LMC \cls using Str\"omgren \phot and high resolution spectroscopy. 
Vallenari {\it et al.} (1991) find a \met of --0.6 for the IAC NGC 2164. As for the LMC field,
there are also several recent indications that we may need to revise \met estimates downwards.
Geha {\it et al.} (1998) find very low \mets (\ltsim --1) are required to fit the giant branch 
colors in their HST fields. And Alves {\it et al.} (1998) analyze MACHO project \phot for 
$\sim 10^7$ stars in the bar and find a two-component model best fits the observations, 
with the intermediate age component having a \met of $\sim -0.7$.

Thus, we conclude that intermediate age  LMC clusters and fields, especially in the outer disk,
may be   more metal-poor than 
previously generally
regarded.
More spectroscopic and photometric
studies with a large number of giants per cluster for a larger sample of IACs
would help clarify this question.

\section{Dual clumps}                           

The fields of SL388 and SL509 (Fig.4), located  together in the northern part of the LMC (Fig.2),
present a unique feature: a populated secondary clump $\approx$0.45 mag fainter than 
the prominent main  clump seen in all other fields.  This fainter clump  in the CMD is  also
slightly bluer, which suggests a lower metallicity and/or older age.
The lower clump stars are found uniformly across both fields.
This feature might be interpreted as
a depth structure consisting of   a layer of stars located approximately 10 Kpc behind the LMC,
at a distance comparable to that of the SMC.  
The fact that the fields of SL509 and 
SL388 are separated by only  $\approx$1$^o$ on the sky sets a lower limit to the 
size of that possible structure of about 1 Kpc. 
An alternative explanation   might be  the presence of 
a dwarf companion to the LMC. In the SMC the occurrence of depth effects is well documented 
both in the field stellar component (Hatzidimitriou \& Hawkins 1989) and  in  velocity space 
for the HI distribution (Mathewson \& Ford 1984). A stellar bridge exists  between the Clouds 
having arisen from their past interaction (Irwin, Demers \& Kunkel 1990), and two 
concentrations in the bridge and one in the extreme of the SMC wing  may
evolve into dwarf galaxies (Bica \& Schmitt 1995). All this evidence suggests that 
depth effects or dwarf galaxies  might also occur elsewhere in the Magellanic System,
as debris from interactions.

Arguments against this interpretation are that, if it is a superposition of two 
similar \pops at
different distances, one would expect at least greater scatter in the principal CMD sequences,
yet this is not the case. More importantly, SL509 itself shows the same dual clump, with the
fainter clump even more pronounced with respect to the brighter clump than seen in the field
CMD, and an equal-area field reinforces this. Also,
there is a hint of a dual clump in the very populous SL769 field, located many degrees away.
More
studies are needed to investigate this intriguing feature.

\section{Conclusions}

By using Washington system  
color-magnitude diagrams of a large sample of IACs             
and their surrounding fields in the LMC outer disk we were able to derive ages 
and metallicities. 
The cluster stellar population is a major component of the field 
where it is embedded, thus sharing its mean  age and metallicity properties, except
in three  cases where the clusters are     $\approx$0.3 dex more metal poor than the
field, suggesting that the chemical enrichment 
was not globally homogeneous in the LMC. Our data is consistent with a scenario in which
local   star formation events generated 
both the clusters and a significant part of their surrounding stellar fields. Thus  during the 
last 1-3 Gyr  the dynamical evolution of the disk has not significantly taken them 
apart, which provides information on 
the  diffusion time-scale  and mixing of stellar generations in the disk. 
The old population (t$\geq$10 Gyr) is detected in most fields as a small  concentration
of stars on the horizontal branch blueward and faintward of the prominent clump. 

 The unique cluster
ESO121-SC03 at $\approx$9 Gyr  
has a  surrounding field which shares  the same stellar \pop.
No other field so far studied is dominated by such an old population.
One possibility would be that a field and cluster population coupling might last that
long. Alternatively, we might be dealing with a building block recently accreted by the LMC in the
form of a dwarf galaxy. 

One IAC cluster (OHSC37) is so far from the LMC body that no surrounding LMC field is
detected.  The present observations suggest that the LMC stellar disk extends
out to between $\sim 11-13$ kpc in deprojected radius.
In the northern part of the LMC outer disk
the fields of SL388 and SL509 present evidence of a depth effect
with a secondary component    located behind the LMC disk at a distance 
comparable to that of the SMC. 
A background layer of stars  in the LMC was possibly detected, and its size  
is at least $\approx$1 kpc, comparable to that of a dwarf galaxy.  The peculiar
location of the cluster  OHSC37 and the depth effect 
in the SL\,388 and SL\,509 fields  might be explained as  debris  from   previous
interactions of the LMC with  the Galaxy and/or the SMC.

The average metallicity derived for the present outer disk IACs     
is $<$[Fe/H]$>$$=-0.71\pm0.17$ and for their surrounding fields 
$<$[Fe/H]$>$$=-0.61\pm0.11$. 
A few clusters stand out in the age--metallicity relation in the sense that they are 
intermediate age clusters at [Fe/H]$\approx -1$.
The outer LMC disk (clusters and fields),
at least of this age range (1-3 Gyrs) seem to be  more metal-poor than 
previously generally
regarded.

The authors would like to thank Cristina Torres for providing data critical to the 
\met calibration in advance of publication. E. Geisler, as always, was an inspiration.
This research is
supported in part by NASA through grant No. GO-06810.01-95A (to DG) from the Space
Telescope Science
Institute, which is operated by the Association of Universities for
Research in Astronomy, Inc., under NASA contract NAS5-26555. This work was partially
supported by the Brazilian institutions CNPq and FINEP, the Argentine institutions
CONICET and CONICOR, and the VITAE and Antorcha foundations.

\newpage

\centerline{\bf Figure Captions}
\begin{figure}
\figcaption{$T_1$ (R)  frame of IC2134 (upper right  cluster) and SL 451. 
The field  is  $13.6\arcmin$ on a  side. North is up and east is to the left.}
\end{figure}

\begin{figure}
\figcaption{Spatial distribution of the  sample. The LMC bar is 
schematically represented by a line for comparison purposes. 
Clusters in the outer disk are shown as triangles.
The fields of SL388
and SL509 (filled triangles) present an extra clump in the CMD suggesting a depth 
structure (see Section 6). The field of SL769 (square) belongs to the inner disk.}
\end{figure}

\begin{figure}
\figcaption{Washington T$_1$ vs. (C-T$_1$) CMDs of the  star clusters. 
Extraction radius in pixels is given in each panel. a. SL8 b. SL126 c. SL262 d. SL388 e. 
IC2134 f. SL451 g. SL509 h. SL817 i. ESO121SC03 j. SL842 k. SL862 l. OHSC33 m. OHSC37 n. SL769}
\end{figure}

\begin{figure}
\figcaption{Washington T$_1$ vs. (C-T$_1$) CMDs of the  surrounding fields, 
excluding areas of radius  two times that of the cluster.  a. SL8 b. SL126 c. SL262 
d. SL388 e. 
IC2134 f. SL509 g. SL817 h. ESO121SC03 i. SL842 j. SL862 k. OHSC33 l. OHSC37 m. SL769}
\end{figure}

\begin{figure}
\figcaption{Metallicity derived from the color of the giant branch in the Washington system
compared to standard giant branches vs. standard \met. Plus signs are LMC \cls, squares are
Galactic open \cls. The solid line shows perfect correlation; the dashed line indicates the
mean relation we find: $[Fe/H]_{STD}=[Fe/H]_{SGB}+0.45$.}
\end{figure}

\begin{figure}
\figcaption{Metallicity derivation for the IAC SL862. The \cl has been placed in the 
absolute \T1 magnitude -- dereddened \CT1 color plane assuming a true distance modulus of
18.5 and a reddening of E(B-V)=0.09. The standard giant branches are those of 
NGC 6752
(offset [Fe/H]=--1.1) and NGC 1851 (offset [Fe/H]=--0.75).}
\end{figure}

\begin{figure}
\figcaption{Metallicity derivation for the field of SL862. Caption as for Figure 6, with
the addition of the standard giant branch for 47 Tuc (offset [Fe/H]=--0.3).}
\end{figure}

\begin{figure}
\figcaption{Metallicity derivation for the old \cl ESO121-SC03.
Caption as for Figure 7, except that the standard, not offset, \mets are shown.}
\end{figure}

\begin{figure}
\figcaption{Chemical evolution of the LMC outer disk: open triangles refer to fields,
filled triangles to clusters. Lines link cluster and the respective surrounding field. 
For comparison purposes we show some LMC globular clusters
(filled circles) and an LMC inner disk cluster (open square). The age for the 
field corresponding to SL126 located at (9.4, -0.5) is a lower limit (Table 3).
The curve shows the mean age-metallicity relation derived by Olszewski {\it et al.} (1991) for 
IACs.}
\end{figure}

\begin{figure}
\figcaption{Metallicity vs. deprojected radius for LMC outer \cls (squares) and fields (plus
signs). The point at the bottom right is ESO121-SC03.}
\end{figure}

\end{document}